\documentclass[twocolumn]{aastex6}
\pdfoutput=1 
\usepackage{amsmath,amstext}
\usepackage[T1]{fontenc}
\usepackage{apjfonts} 
\usepackage[figure,figure*]{hypcap}

\shorttitle{A Monster CME Obscuring A Demon Star Flare}
\shortauthors{S.P. Moschou et al.}

\begin{document}

\title{A Monster CME Obscuring A Demon Star Flare}

\author{Sofia-Paraskevi Moschou}
\author{Jeremy J. Drake}
\affiliation{Smithsonian Astrophysical Observatory, 60 Garden Street, Cambridge MA02138, USA}

\author{Ofer Cohen}
\affiliation{Lowell Center for Space Science and Technology, University of Massachusetts, Lowell, Massachusetts, USA}
\author{Julian D. Alvarado-Gomez and Cecilia Garraffo}
\affiliation{Smithsonian Astrophysical Observatory, 60 Garden Street, Cambridge MA02138, USA}

\begin{abstract}
We explore the scenario of a Coronal Mass Ejection (CME) being the cause of the observed continuous X-ray absorption of the August 30 1997 superflare on the eclipsing binary Algol (the {\it Demon Star}). The temporal decay of the absorption is consistent with absorption by a CME undergoing self-similar evolution with uniform expansion velocity.  
We investigate the kinematic and energetic properties of the CME using the ice-cream cone model for its three-dimensional structure in combination with the observed profile of the hydrogen column density decline with time. Different physically justified length scales were used that allowed us to estimate lower and upper limits of the possible CME characteristics. Further consideration of the maximum available magnetic energy in starspots leads us to quantify its mass as likely lying in the range $2\times 10^{21}$--$2\times 10^{22}$~g and kinetic energy in the range $7\times 10^{35}$--$3\times 10^{38}$ erg. The results are in reasonable agreement with extrapolated relations between flare X-ray fluence and CME mass and kinetic energy derived for solar CMEs. 
\end{abstract}


\section{Introduction}
\label{S:1}

Since the first space-based coronagraphic observations of solar coronal mass ejections (CMEs) in the 1970s \citep{Tousey:73}, the study of CME properties has been pursued with some vigor due to their implications for space weather and potential impact on terrestrial life \citep{Kahler.etal:01,Zhang.etal:07,Webb.etal:09,Yashiro2009,Cane.etal:10,Vourlidas2011,Cliver.Dietrich:13,Reames:13,Gopalswamy:16}. The growing realisation that exoplanets are extremely common in the universe and that their host star CMEs might influence their atmospheric evolution \citep[e.g.][]{Khodachenko.etal:07,Khodachenko.etal:07b} has raised the question of the nature of CMEs on other stars \citep[e.g.][]{Kay.etal:16}.  CMEs on the Sun are associated with flares, and it has also been pointed out that the winds of magnetically active stars with much more vigorous flare activity than the Sun could be dominated by CMEs, with potentially important implications regarding the large amount of energy that might be involved \citep{Drake.etal:13} . 

Unfortunately, the technological level of current instrumentation does not yet allow for direct observations of stellar CMEs. In order to attempt to study them we need to recruit indirect methods and techniques. One such technique that offers perhaps more promise for large stellar CMEs than solar ones is the absorption of the underlying corona by CME material.  While absorption is seen in CME filaments on the Sun \citep[e.g.][]{Subramanian.Dere:01,Kundu.etal:04,Jiang.etal:06,Vemareddy.etal:12}, there is generally too little material present in the CME itself to cause large-scale absorption.  If CMEs associated with the much more energetic flares seen on stars are commensurately more massive, as solar flare--CME relations indicate might be the case \citep[e.g.][]{Yashiro2009,Drake.etal:13}, then CME absorption signatures could be a feasible means of their detection. Indeed, several examples of transient increases in X-ray absorption or obscuration in stellar observations have been identified as potentially having been caused by CMEs or prominences  \citep{Haisch.etal:83,Ottmann.Schmitt:96,Tsuboi.etal:98,Franciosini.etal:01,Pandey.Singh:12}.

One of the most energetic X-ray flares ever observed on a star was the 1997 August 30 event on the 
bright and nearby (28.5~pc) prototypical eclipsing binary system, Algol (also known as the Demon Star due to its association to the mythological monster Medusa). The flare was observed by BeppoSAX and analysis by \citet{Favata1999} found the total X-ray fluence in the 0.1--10~keV band to be $1.4\times10^{37}\;\mathrm{erg}$, or approximately $1 \times10^{37}\;\mathrm{erg}$ in the 1--8~\AA\ GOES band. To place this in the context of solar flares, the largest X-ray fluence in the compilation of flares associated with CMEs by \citet{Yashiro2009} is $6.5\times 10^{30}\;\mathrm{erg}$, while the great Carrington Event of 1859 has been estimated to have had a soft X-ray fluence of $1.8\times 10^{31}$~erg and a total radiated energy of $5\times 10^{32}$~erg \citep{Cliver.Dietrich:13}. The 1997 August 30 Algol flare was, staggeringly, about 10,000 times more energetic than this.  The flare was eclipsed by the primary star, enabling \citet{Schmitt1999Natur} to estimate both its location and size.

One other feature of the flare was a large increase in absorption at the flare onset that gradually decayed back to the interstellar medium value. \cite{Favata1999} suggested a coronal mass ejection as the source of the absorption.  In fact, this event arguably presents the best characterized observational evidence of a CME on a star other than the Sun and a valuable opportunity to explore stellar CME properties. Here, we seek to exploit this opportunity and investigate the CME scenario using the parameterized geometrical ``ice cream cone" model developed to analyse solar CMEs by \cite{Howard82} and later by \cite{Xie2004}. 

We first reprise the details of the 1997 August 30 event in Section~\ref{s:flare}, and then describe briefly in Section~\ref{s:icecream} the ice cream cone model that we use in Section~\ref{sec:analysis} to analyse the data.
Using the observed flare and inferred CME characteristics we then explore the mass and kinetic energy implications in the context of solar and stellar flares and CMEs in Section~\ref{s:discuss}.

\section{The 1997 August 30 Algol Flare}
\label{s:flare}

Algol is the prototype of the Algol-type binaries---short-period, eclipsing systems comprising an early-type primary and a late-type main-sequence or subgiant secondary.  These systems have generally undergone a period of mass
transfer during which material from the initially more
massive present day late-type star has been accreted by its initially less massive present day early-type companion.  Algol itself is a B8~V primary with a K2~IV secondary that has lost about half of its original mass to the present-day primary \citep{Drake:03}. The stellar parameters are $R_A = 2.90 R_\odot$, $M_A = 3.7 M_\odot$, $R_B = 3.5R_\odot$ and $M_B = 0.81M_\odot$, with an orbital period of 2.87 days and inclination $i = 81.4$~deg \citep{Richards:93}. The orbital separation is $14.14 R_\odot$. A fainter tertiary component with late A or early F spectral type is also present in a much wider 1.86~yr orbit \citep{Bachmann.Hershey:75}.

Tidal spin-orbit coupling tends to lock the rotation period of Algol components to that of the orbit, such that both stars of the binary are rapid rotators. The rapid rotation excites magnetic dynamo action in the convection zone of the late-type star that is manifest in the form of chromospheric and coronal emission \citep[e.g.][]{Drake.etal:89,Singh.etal:95}.  
Radio and X-ray activity levels of Algol systems are, not surprisingly, quite 
similar to, but often slightly lower than, those of the 
short period late-type RS~CVn-type binaries \citep{Singh.etal:96,Sarna.etal:98}.  
The B8~V component of Algol was confirmed as being essentially X-ray dark though Doppler analysis of high resolution {\it Chandra} X-ray spectra by \citet{Chung.etal:04}, such that all the observed X-ray emission is from the K2 subgiant.
The brightness of Algol at X-ray wavelengths has rendered it a popular target for X-ray satellites \citep[see, e.g., the summaries of][and references therein]{Favata1999,Chung.etal:04}.

\begin{figure}[htbp]
\begin{center}
\includegraphics[width=\columnwidth]{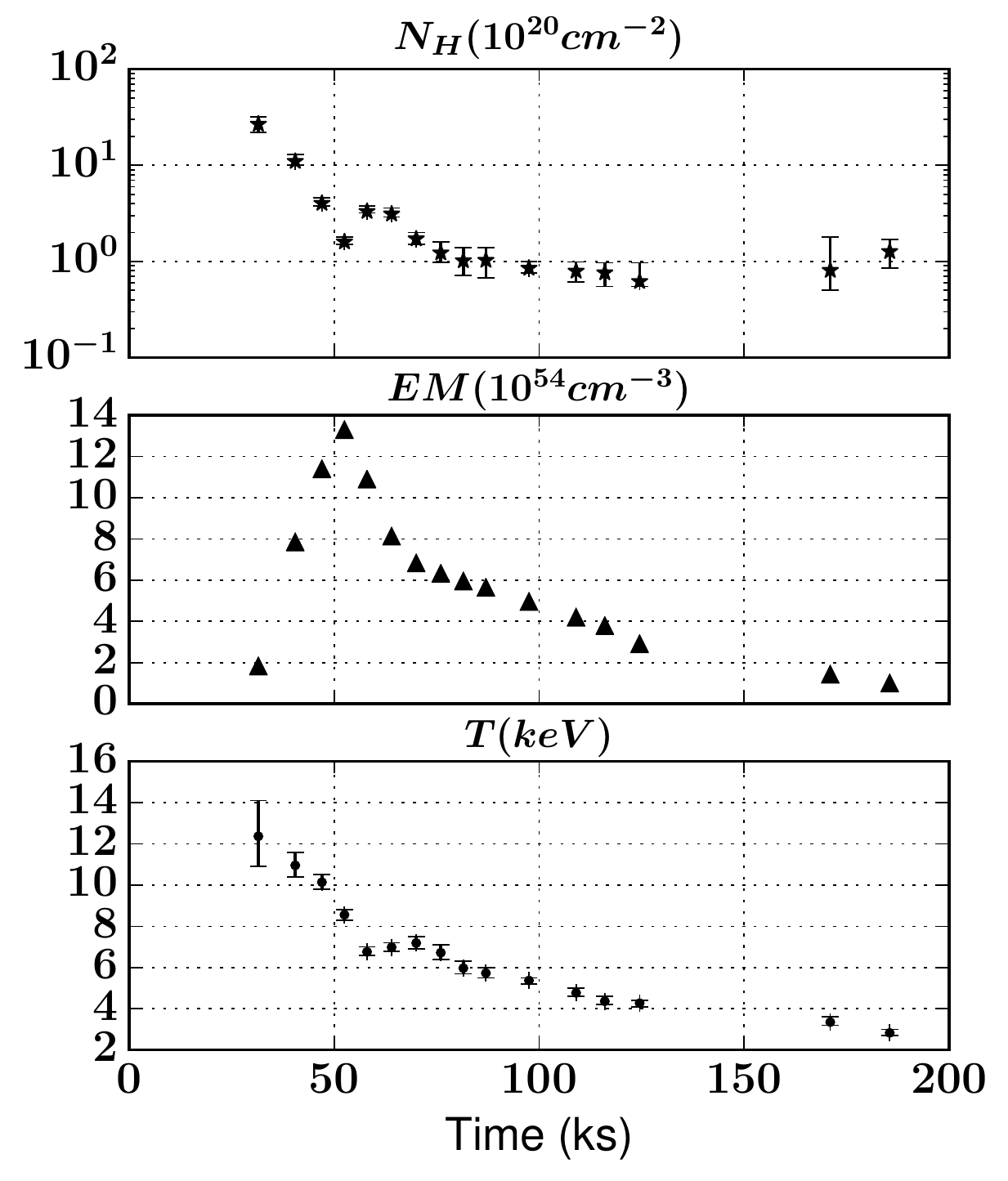}
\caption{Temporal profiles of the column density, emission measure and temperature as estimated by the best-fit in \cite{Favata1999}.}
\label{fig:favdat}
\end{center}
\end{figure}

BeppoSAX \citep{Boella.etal:97} observed Algol over a period of about 240~ks covering almost a full orbit starting on 1997 August 30 03:04 UT \citep[see][for further details]{Favata1999}. During the observation an enormous flare was observed whose decay had not fully reached quiescent levels by the end of the exposure.  Detailed parameter estimation using optically-thin collision-dominated radiative loss models was performed by \citet{Favata1999}, who derived time-dependent plasma properties throughout the observation, including plasma metallicity, emission measure, temperature and intervening absorption.  The temporal variations of the last three quantities for the flare duration are reproduced in Figure~\ref{fig:favdat}. Of special note here is the large increase in absorption coinciding with flare onset. 
Both the absorbing hydrogen column density and the temperature decrease monotonically with time, while the emission measure starts from background values, to reach a peak at 50 ks and return close to background values after 200~ks from the beginning of the measurements.

Crucially, the flare was totally eclipsed by the primary star, which enabled \citet{Schmitt1999Natur} to determine that the plasma was confined to Algol~B.  By modelling the light curve, they concluded the flare occurred near the south pole, reached a maximum height of 0.6 stellar radii, and that continuous heating would have been required, similar to two ribbon flares on the Sun except involving orders of magnitude more energy. \citet{Sanz-Forcada.etal:07} found that the location interpretation of \citet{Schmitt1999Natur} was not necessarily unique, although this is not important for the purposes of our analysis.
While solar flares are rarely observed in polar regions \citep{Joshi.etal:10}, rapidly rotating stars are observed to have large polar spots \citep{Schuessler1992,Strassmeier:09}, which could be the origin of the observed superflare.

\citet{Favata1999} determined the flare onset to be at the start of their interval 2, or at $t=26.3$~ks, in which the flaring site was already obscured by absorption. 
In our subsequent analysis, we adopt the hydrogen absorption as a function of time derived by \citet{Favata1999} in order to examine the likely parameters of the CME thought to be responsible.  

\section{The Ice Cream Cone CME Model}
\label{s:icecream}

First introduced by \cite{Howard82}, the ice cream cone CME model is a geometric model, the parameters of which can readily be determined through coronagraph observations. It was the first use of a 3D bubble-like topology, instead of the 2D loop-like models used until that time. 
Later on, \cite{Zhao2002} fully established the cone model for halo CMEs, making three concrete assumptions: a) the CME source location is at the center of the solar disk close to the associated active region (AR) 
surface area;
b) it has a radial bulk velocity; and
c) constant angular width throughout its propagation.

\cite{Xie2004} further improved the cone model by providing analytic relations to derive the actual orientation, angular width and speed of halo CMEs from geometric arguments based on coronagraph observations, also assuming isotropic expansion of the CME. They noted that CME speeds vary greatly from 100 up to $2500\; \mathrm{km\ s^{-1}}$, with the slower ones at heliocentric distances of a few solar radii appearing to accelerate and then keep a constant speed, while the faster ones appear to decelerate. 



\begin{figure}[htbp]
\begin{center}
	\includegraphics[clip, trim=11cm 5.5cm 11cm 0cm, width=\columnwidth]{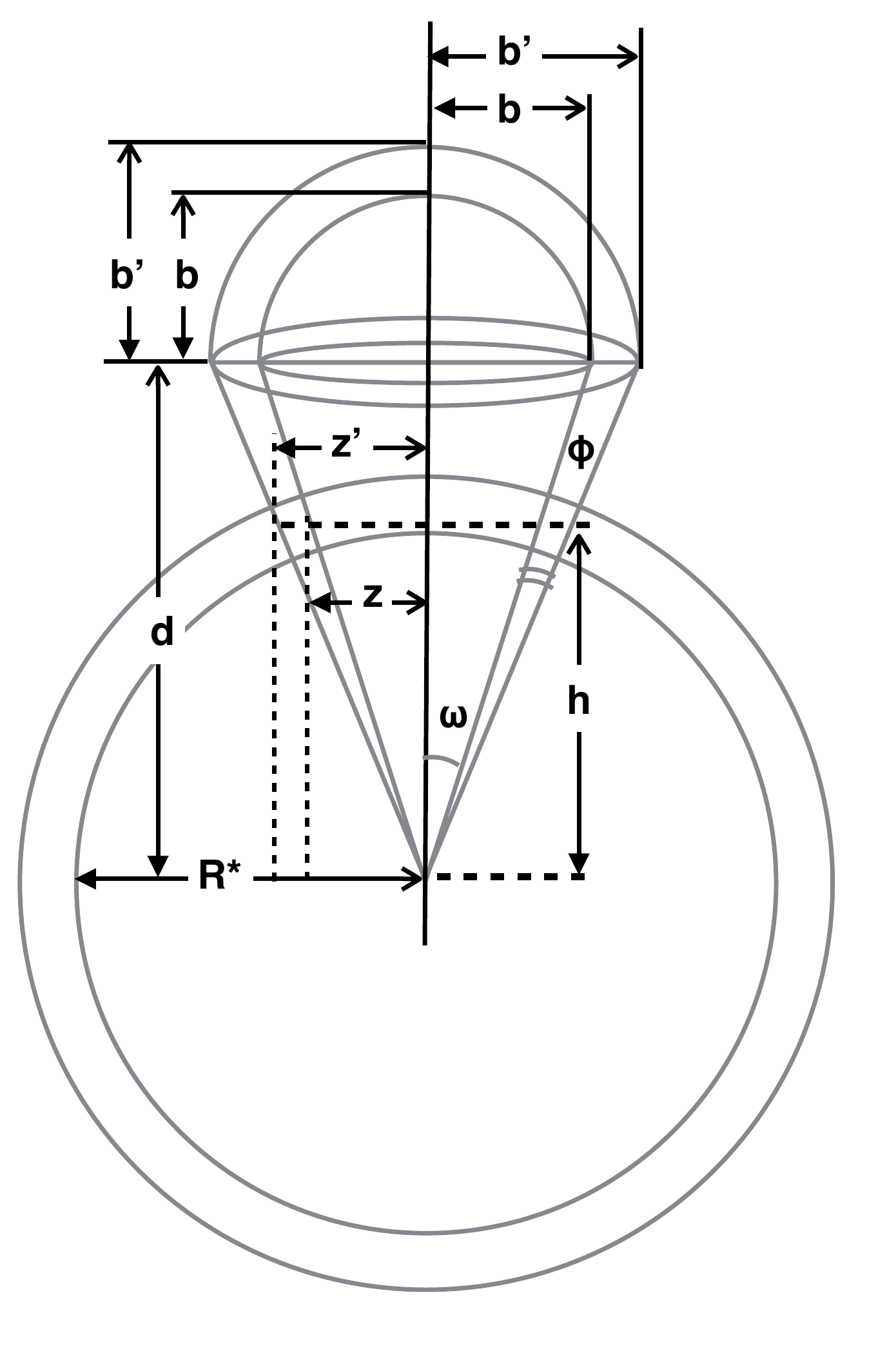}
\caption{Sketch showing the CME structure according to the ice-cream cone model, following \citet{Fisher1984}.}
\label{fig:cme}
\end{center}
\end{figure}

\cite{Fisher1984} formalized the three-dimensional structure of a CME under the cone model approach into two shells, a truncated conical and a hemispherical one. The volumes and surface areas of each shell are given by
\begin{eqnarray}
\label{eq:vcone}
V_{cone}=\frac{d}{3} \pi (b'^2-b^2)-\frac{h}{3} \pi (z'^2-z^2) {,}\\
\label{eq:vhemi}
V_{hemi}=\frac{4}{3}(b'^3-b^3) {,} 
\end{eqnarray}
where d is the height of the cone, so that $h=R_{Algol,B} \cos(\omega+\phi)$.  We make the additional assumption that the ice cream part of the CME is hemispherical, as illustrated in Figure~\ref{fig:cme}, rather than the more general ellipsoid considered by \citet{Fisher1984}; in their nomenclature we take $\alpha=b'$ and $\alpha'=b$. Solar observations suggest that in a plethora of cases CMEs tend to preserve their global configuration during their evolution and theoretical models have been built based on a self-similar approximation \citep[e.g.][]{Low1987, GibsonLow1998}. 

Applying the cone model from the stellar center outwards, we have that the total distance from the stellar center to the CME front is given by $d(t)+b'(t)=S(t)+R_{Algol,B}$ and thus the cone height from the stellar center can be obtain by $d(t)=(S(t)+R_{Algol,B})/(\tan(\omega+\phi)+1)$. Note that the ratio of $d$ to $b'$ is fixed by the opening angle of the cone.
If the opening angle of the cone is $2\omega$, then the outer and inner radii of the ice cream part of the model are $b'=\tan{(\omega+\phi)}d$ and $b=\tan{\omega}d$, respectively, while the cone radii at the stellar surface height are $z'=\tan{(\omega+\phi)}h$, $z=\tan{\omega}h$ for the outer and inner parts of the shell respectively, through simple trigonometric arguments (Figure~\ref{fig:cme}). 
We can then calculate the hemispherical radius $b'$ for the ice cream part, once we assume a conical opening angle. 

The opening angle $\phi$ of each leg of the CME can be calculated through geometric arguments as $\phi=\tan^{-1}\left(b'/d\right)-\omega$.  \cite{Lepping.etal:90} conclude that magnetic clouds have thickness of about 0.2-0.4AU. \cite{Mulligan.Russell:01} fit the parameters of two previously observed CMEs using a flux rope model. Using \cite{Mulligan.Russell:01} values for the CME cone opening and the \cite{Lepping.etal:90} for the CME thickness, we constraint $\phi$ between [$2^\circ,16^\circ$].

\section{Analysis} \label{sec:analysis}

\subsection{CME propagation direction}
\label{subsec:direction}

\begin{figure}[htbp]
\begin{center}
\includegraphics[clip, trim=0cm 3cm 0cm 3cm, width=\columnwidth]{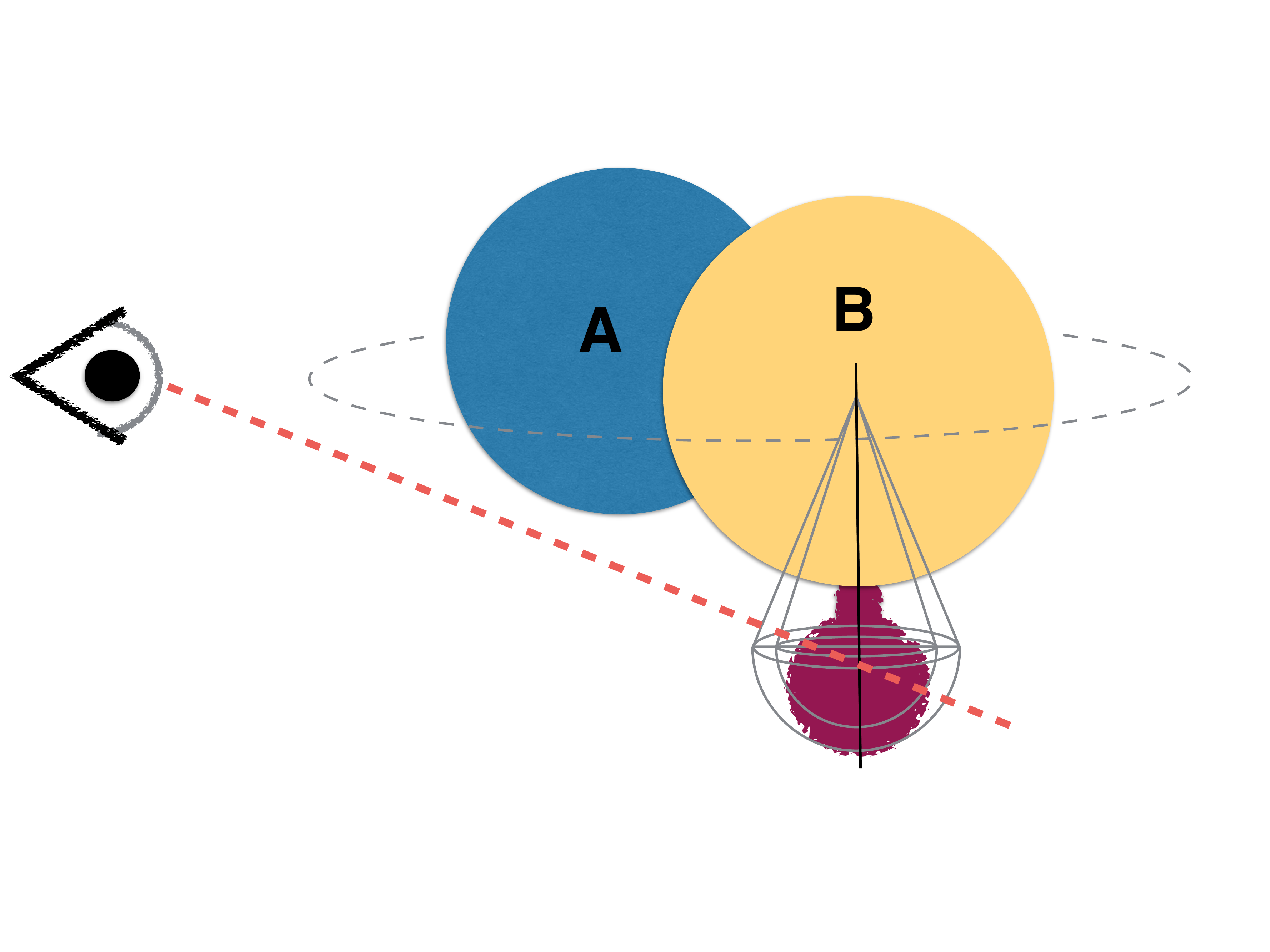}
\includegraphics[clip, trim=0cm 3cm 0cm 3cm, width=\columnwidth]{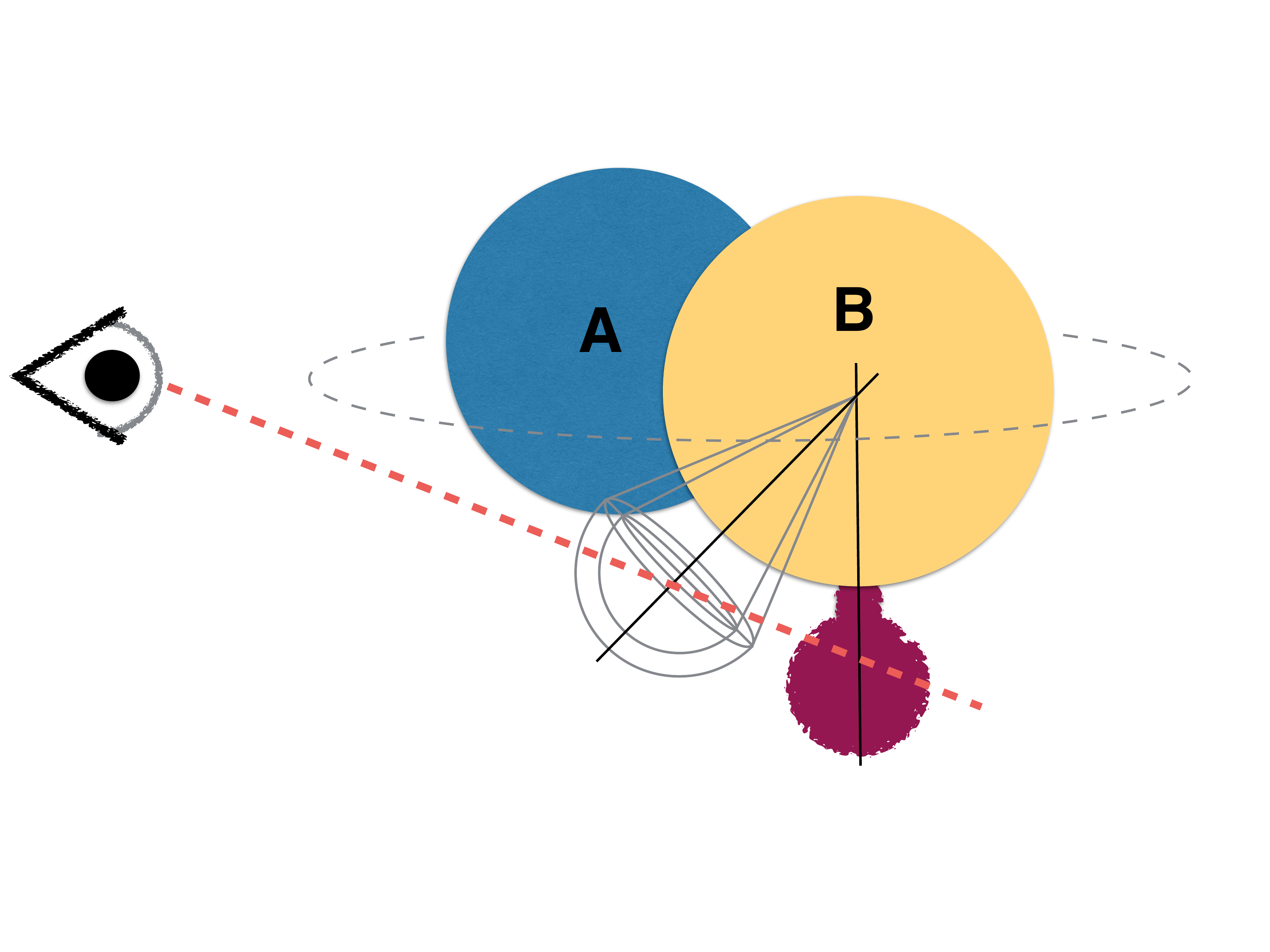}
\caption{Sketch showing the Algol binary system and the ice cream cone model with the possible angular CME ejection range. Top panel: CME propagates directly out of the South pole, Bottom panel: CME propagates at a direction forming a $45^\circ$ angle with the astrographic axis.}
\label{fig:algolcme}
\end{center}
\end{figure}

It is likely that the CME was ejected in the South hemisphere of Algol B, and probably close to the flaring site that was inferred to be near the South pole by \citet{Schmitt1999Natur}. \citet{Sanz-Forcada.etal:07} argued that other flare locations are possible, though the particulars are not important for our analysis and for the purposes of clarity we assume hereafter the configuration deduced by \citet{Schmitt1999Natur}.  According to solar CME data compiled by \cite{Yashiro.etal:08} and \cite{Aarnio2011SoPh}, the separation angle between CMEs and associated flares is $0^\circ\pm45^\circ$, which means that our CME could be ejected directly out of the South pole or with its propagation symmetry axis forming an angle of up to $45^\circ$ with respect to that of the flare, as demonstrated in Figure~\ref{fig:algolcme}. 
%

In general, the line of sight can go through the CME mass in three different main directions a) looking through the ice-cream and one side of the cone shell, b) looking through one side of the cone shell and the edge of the cone and ice-cream, or c) looking through both sides of the cone shell volume. Each case will imply a different path length for the integral of the column density, as will be discussed in Subsection \ref{subsec:mass}. 

\subsection{CME speed}
\label{s:speed}

The CME will propagate and expand outwards in one of the possible directions discussed in Subsection \ref{subsec:direction}.
 The general propagation profile for a CME is an initial acceleration phase followed by a cruising phase at quasi-constant velocity. While fast CMEs (e.g.\ $V>1000$ km $\mathrm{s^{-1}}$) can show deceleration in the LASCO field of view (2.5--30 $R_\odot$) due to the interaction with the solar wind \citep[e.g.][]{Manoharan:06}, we shall see below that the hydrogen column density evolution indicates a constant velocity cruising phase such that the CME is not ``fast'' and subject to significant deceleration within the Algol~B wind. In this case, the plasma travels a distance $S$ with time $t$ equal to

\begin{eqnarray}
\label{eq:genacc}
S_1(t)=\frac{\alpha t^2}{2} {,}\\
S_2(t)=u_{max}(t-t_1) {,}
\end{eqnarray}
 so that the total distance covered by the CME in time t is 
 \begin{equation}
   S_{tot}(t)=\begin{cases}
     S_1(t), & \text{if $t\leq t_1$},\\
    S_1(t)+S_2(t), & \text{if $t>t_1$}.
   \end{cases}
\end{equation}
Here, $\alpha$ is the acceleration and $u_{max}$ the terminal velocity of the CME.

The column density within the CME shell is a measure of the number of absorbers per unit surface area in our line of sight. For an initial examination of the data, if we assume that the CME is a piece of a spherical shell and the number of absorbers on the sphere with radius $\mathrm{r(t)}=S(t)$ is $\mathrm{N_a(t_0)}$ at an initial moment $\mathrm{t_0}$ then, due to conservation of mass and extending this to the number of absorbers, we have
\begin{eqnarray}
\label{eq:sphere}
N_H(t)=\frac{N_a}{4\pi r^2(t)} {,}
\end{eqnarray}
which means that the hydrogen column density scales with the inverse square power of radius $N_H(t)\propto r^{-2}$. The column density of a CME of an artitrary shape that expands self-similarly would have the same radial dependence, as the CME shell thickness would be of size $r$, with a number density density decreasing as $r^{-3}$ and the line-of-sight shell thickness being of size $r$, which would lead to a column density $\propto r^{-2}$. In other words, the CME expands outwards while interacting with the surrounding stellar wind in a spherical diverging geometry. We can extract information regarding the CME kinematic properties (distance, velocity, acceleration) from the column density evolution with time. 

\begin{figure}[htbp]
\begin{center}
\includegraphics[width=\columnwidth]{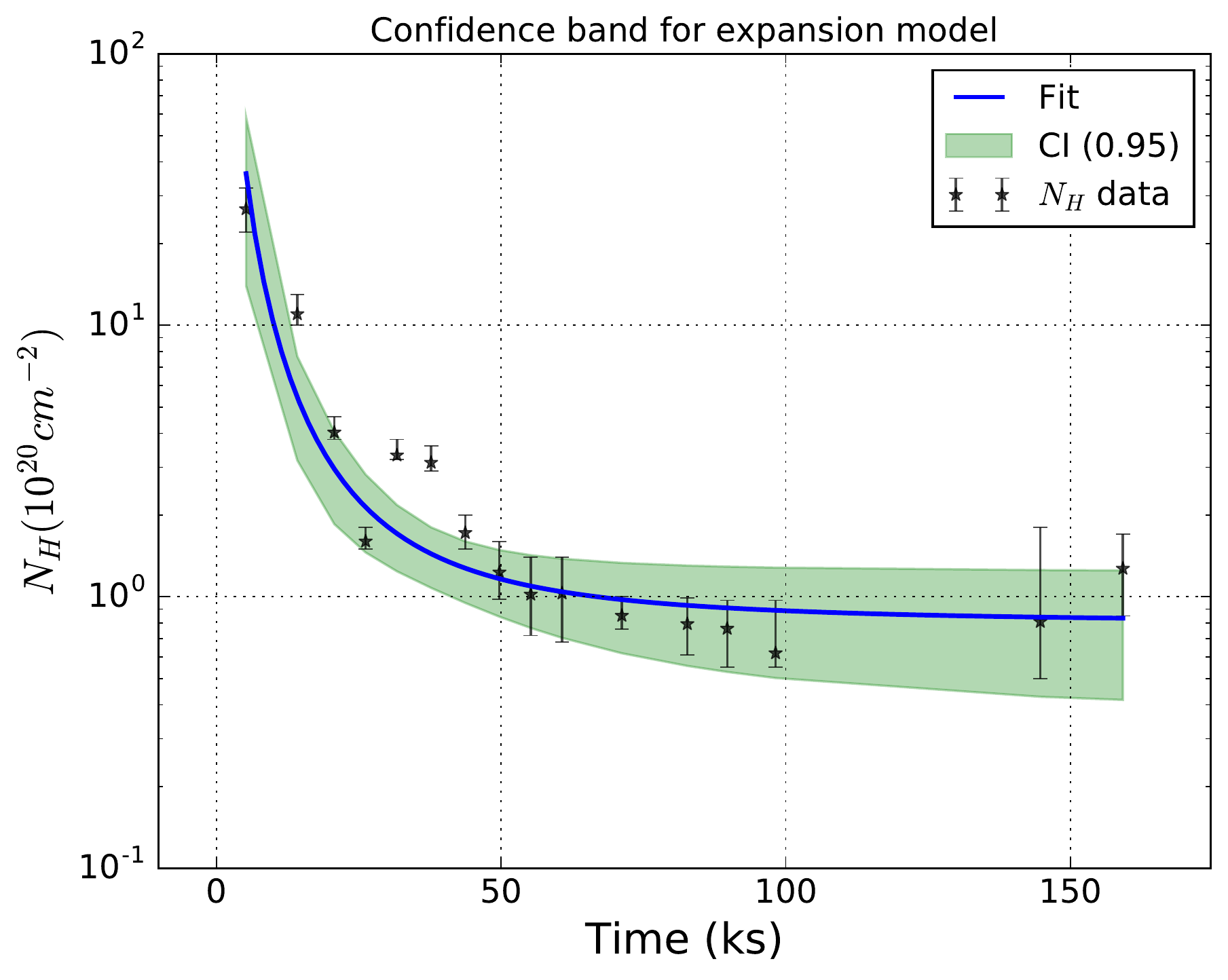}
\caption{Column density temporal profile assuming that the CME is ejected at time $t_0=26.3$ ks.}
\label{fig:nhtime}
\end{center}
\end{figure}

The total column density observed consists of the column density of the CME in addition to the column density of the interstellar medium so that the observed column density is $N_{H,tot}(t)=N_H(t)+N_I$. 
From Equations \ref{eq:genacc} and \ref{eq:sphere}, and by fitting the column density temporal variation, $N_{H,tot}(t)$, to a power law, we can characterize the type of motion that the CME is performing, i.e., if there is any acceleration or deceleration taking place. 
Figure~\ref{fig:nhtime} shows the observed column density temporal profile together with associated errors from \citet{Favata1999}, overplotted with a least squares fit of a power law plus a constant and the 95\%\ confidence interval obtained using the Python based {\it Kapteyn} package \citep{KapteynPackage:15}. 
In units of $10^{20}$~cm$^{-2}$ the fitting gives 
\begin{equation}
N_H(t)=(1000\pm 200)t^{-(2.0\pm1.4)}+(0.8\pm 0.1), 
\label{eq:nh_vs_t}
\end{equation}
such that 
$N_I=(0.8\pm0.1)\times 10^{20}\mathrm{cm^{-2}}$.  The interstellar absorption is in good agreement with the value $(0.9\pm 0.4)\times 10^{20}\; \mathrm{cm^{-2}}$ obtained by \citet{Favata1999} by modelling the pre-flare and secondary eclipse data. 
There is, then, a decay of the local hydrogen column density with time $N_H\propto t^{-2}$, albeit with some uncertainty, which suggests that the acceleration phase of the CME 
has stopped and it has entered the propagation phase, i.e.\ it travels with approximately constant speed.  This conclusion should be valid regardless of the exact shape of the CME provided it expands in a self-similar fashion.  While the CME scenario is not the only possible explanation for the additional absorption of the flare, the agreement of its temporal decline with simple expansion at constant velocity adds considerable weight to this interpretation. 

Assuming we are observing the quasi-constant velocity phase of the CME, and based on the best-fit parameters to the $N_H$ decay profile, we can estimate the CME velocity given a length scale for a particular time. \citet{Favata1999} found the $N_H$ increase to coincide with the flare rise, such that the CME covered the flare beginning at time $t=26.3$.
A natural choice for the CME length scale at this time is then one that obscures the flaring region at the initial stages of the eruption.
\cite{Favata1999} have estimated that the magnetic loop associated with the flare should have a length of about $10^{11}\mathrm{cm}$, which is about $S(t_1)=2R_\odot$.  This actually represents the lower limit to the size of the CME at this time. We will in addition to that, investigate a dynamic length scale, as estimated by models and observations that have studied the force balance mechanism that drives CMEs in the solar regime \citep{Vrsnak2004,Zic2015}. 
After an acceleration phase close to the Sun due to driving by Lorentz forces, the so-called aerodynamic drag force \citep{Cargill2004} starts to dominate at a few solar radii, 
 as demonstrated in \cite{Zic2015}. Inspired by \cite{Zic2015}, who found a dynamic length scale for the acceleration phase of solar CMEs of $S_1(t_1)\sim 15R_\odot$, we adopt a dynamical length scale for the Algol CME of $S_1(t_1)=15R_{Algol,B}$. 

From Equation (\ref{eq:nh_vs_t}) and the fit corresponding to Figure \ref{fig:nhtime}, we estimate the decaying time for the absorption, i.e.\ the time the observed column density $N_H$ takes to reach one quarter of its initial peak value $N_H(\tau_{1/4})=N_H(t_1)/4$, at $\tau_{1/4}=5.6$ ks having traveled for one length scale with constant speed $u_{max}$, where $N_H(t_1)$ the column density of the first observational point in Figure \ref{fig:nhtime}.
Then, we calculate the speed of the CME using the decay time and a) the flare and b) the dynamic length scales, as they were determined above, getting $2.5\times 10^7\mathrm{cm\ s^{-1}}$ and $6.6\times 10^8\mathrm{cm\ s^{-1}}$, respectively, since 
\begin{eqnarray*}
N_H\approx\frac{N_a}{4\pi(S_1+u_{max} t)^2} {,}
\end{eqnarray*}
with $N_a=4\pi S_1^2N_H(t_1)$. The speeds differ by a factor of 26, similar to the length scale differences, and bracket almost the entire velocity range observed in solar CMEs \citep[e.g.][]{Yashiro.etal:04}.

\subsection{CME mass and kinetic energy}
\label{subsec:mass}

Now, by assuming a geometric model for the mass distribution of the CME in three-dimensional space, we can estimate the CME mass and kinetic energy. Here, we apply the ice cream cone model from Section~\ref{s:icecream} \citep{Howard82,Fisher1984,Zhao2002,Xie2004} applied extensively to solar events to further investigate the CME scenario for the Algol flare.


We have for the column density, neglecting the non-uniform thickness of the cone segment, 
\begin{eqnarray}
\label{eq:cone_int}
N_H(t)=\int \frac{N_a}{V_{cone}+V_{hemi}} \; dS
\end{eqnarray}
where $dS$ is the path length through the gas within the ice cream cone. We do not know if the flare is observed inside the CME volume, or if it was behind the CME.  Since the opening angle of solar CMEs is strongly correlated with CME energy and reaches 180$^\circ$ for the main CME body \citep[e.g.][]{Kahler:89,Yashiro2009,Aarnio2011SoPh}, the former scenario is more likely. The difference is a factor of order 2 in the resulting column density, which we neglect here.


The mass of the CME can thus be calculated from
\begin{eqnarray}
\label{eq:mass}
M=\frac{V_{cone}+V_{hemisphere}}{b'-b}N_H\mu {,}
\end{eqnarray}
where $\mu=1.36 m_p$ is the mass per proton for gas of solar composition (X=0.738, Y=0.249, Z=0.013), and $m_p$ is the proton mass.   For a solar composition plasma, the dominant contribution to the soft X-ray absorption cross-section is from inner-shell ionization of the abundant elements O, C, N and Ne. Our mass estimate then assumes that the absorbing material is not highly ionized.  This is consistent with observations of solar CME material \citep[e.g.][]{Webb:12} and with our estimates of the mass, which we note in Section~\ref{s:cmemass} below is inconsistent with having been derived from highly ionized coronal plasma.





As noted previously, the total column density is the sum of the interstellar medium value and the CME one.
For the former we will use the value obtained by the fit illustrated in Figure \ref{fig:nhtime} (Equation~\ref{eq:nh_vs_t}). Then combining with Equations (\ref{eq:vcone}), (\ref{eq:vhemi}) and (\ref{eq:mass}) for the volume of the CME, we can calculate the CME mass for both length scales, i.e.\ flare and dynamic.
In order to explore the parameter space, we examine a set of six $(\omega+\phi)$ angles from $15^\circ$ to $90^\circ$ in $15^\circ$ steps.  
Fixing the CME thickness at a distance of 1AU to $\sim$ 0.2 AU to match solar observations, as discussed in Section~\ref{s:icecream}, we can determine the thickness of the CME at each length scale considered, since the model assumes a self-similar expansion, which will in turn define the angle $\phi$. In this way, a thickness of $0.4R_\odot$ and $3R_{Algol,B}$ were determined for the flare and dynamic length scales, respectively.
The results for both mass and kinetic energy for the CME are shown in Table \ref{tab:CME}.  

Both the inferred CME mass and energy vary by an order of magnitude according to the opening angle assumed.  Larger opening angles imply larger CMEs and so larger mass and energy values.  However, the biggest influence on the inferred CME parameters is the assumed scale length, which, for the values we have assumed, lead to mass and energy differences of two and five orders of magnitude, respectively.  We will return to this in Section~\ref{s:discuss} below.


 \begin{table}[htbp]
 \centering
 \begin{tabular}{ccccc}
 \hline
 $\omega+\phi$ & $M_{obs} (g)$ &$E_{k,obs} (erg)$ & $M_{dyn} (g)$ &$E_{k,dyn} (erg)$ \\
 \hline
 $15^\circ$ & $3.1\times 10^{20}$ & $9.7\times 10^{34}$ & $3.2\times 10^{22}$ & $6.9\times 10^{39}$\\
 $30^\circ$ & $6.0\times 10^{20}$ & $1.9\times 10^{35}$ &  $7.7\times 10^{22}$ & $1.7\times 10^{40}$ \\
 $45^\circ$ & $9.8\times 10^{20}$ & $3.1\times 10^{35}$ &  $1.2\times 10^{23}$ & $2.6\times 10^{40}$\\
 $60^\circ$ & $1.6\times 10^{21}$ & $4.9\times 10^{35}$ &  $1.7\times 10^{23}$ & $3.7\times 10^{40}$\\
 $75^\circ$ & $2.4\times 10^{21}$ & $7.4\times 10^{35}$ & $2.3\times 10^{23}$ & $5.0\times 10^{40}$ \\
  $90^\circ$ & $3.4\times 10^{21}$ & $1.1\times 10^{36}$ & $3.2\times 10^{23}$ & $6.8\times 10^{40}$\\
 \hline
 \end{tabular}
 \caption{Table with opening ice-cream cone model angles, estimated corresponding CME masses and kinetic energies. The CME thickness for a) the flare and b) the dynamic length scales were chosen as $0.4R_\odot$ and $3R_{Algol,B}$, respectively, to match the observed values from the solar case 1AU of about 0.2 AU.}
 \label{tab:CME}
 \end{table}

\begin{figure}[htbp]
\begin{center}
\includegraphics[width=\columnwidth]{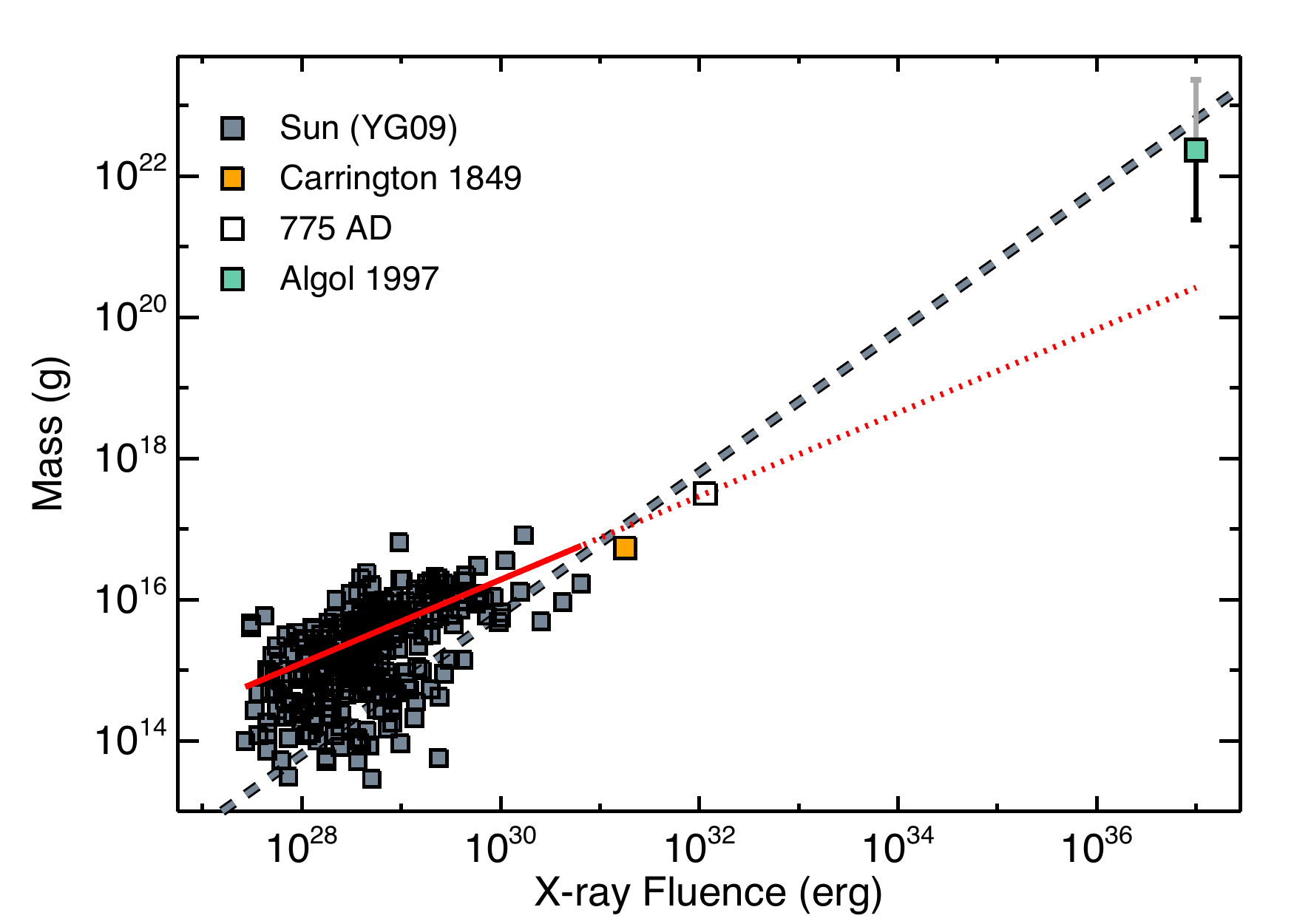}
\includegraphics[width=\columnwidth]{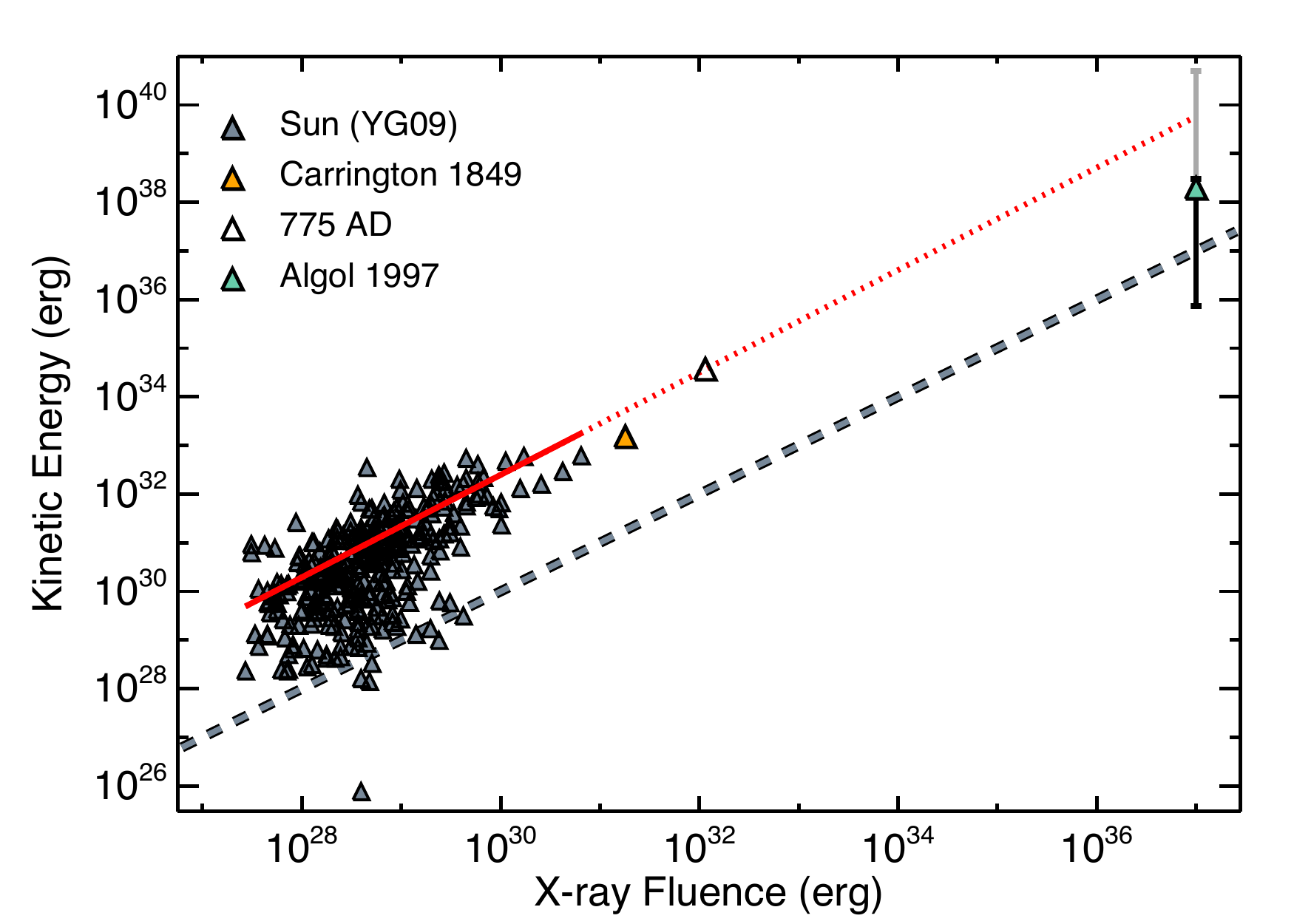}
\caption{Comparison of the derived 1997 April 30 Algol CME properties for opening angle $\omega+\phi=75^\circ$ (see Section~\ref{s:solar}) as a function of X-ray flare fluence in the 1--8\AA\ band with the solar flare--CME compilation of \cite{Yashiro2009}. Top: CME mass. Bottom: CME kinetic energy. Grey extensions to the error bars represent the full range of derived values, while black error bars show the restricted ranges considering the available magnetic energy (see Section~\ref{s:energy}). Red lines represent the mean relations derived by \citet{Drake.etal:13} and their extrapolations. In the top panel, the dashed gray line follows a constant ratio of mass loss to GOES X-ray fluence, $\dot{M}=10^{−10} (L_X /10^{30}) M_\odot$ yr$^{−1}$ . In the lower panel, the dashed gray line represents equivalence of CME kinetic and flare X-ray energies.}
\label{fig:solar}
\end{center}
\end{figure}

\section{Discussion}
\label{s:discuss}

The parameters derived for the Algol CME scenario are quite extreme in the context of solar eruptive events, which is not surprising considering the enormous flare energy involved.  While there are large uncertainties in the derived mass and energy, the results still provide valuable insights into the properties of such events on the most active stars.  We examine some of the implications below.  

\subsection{Comparison with solar CMEs}
\label{s:solar}

Since our derived CME properties have a significant dependence on the opening angle assumed, it is worthwhile to assess what the likely value of this parameter would be for such an event.

\cite{MichaleGopalswamy2003} performed a statistical analysis of CME parameters using the cone model for all the halo CMEs detected with SOHO/LASCO from the end of June 1999 until the end of of 2000. For close to solar maximum conditions corresponding to the 23rd solar cycle, they concluded that the average cone opening angle is $60^\circ$, with the most probable value being $67.5^\circ$, while the average velocity is $1080\;\mathrm{km\; s^{-1}}$ and the most probable one $600\;\mathrm{km\; s^{-1}}$. 
\cite{Aarnio2011SoPh} quantified that 4 out of 5 CMEs that are linked to X-class flares are of halo type, with wider opening angles, and indicated that the most probable CME width is around $155^\circ$, or $\omega\approx 77.5^\circ$.  The statistical analysis in \cite{Yashiro2009} revealed that the CME width is correlated with both the total flux emitted by the associated flare, and with CME kinetic energy, with wider CMEs being more energetic. 

If solar CMEs are a guide to the Algol event, the opening angle of this very energetic CME should be toward the high end of the values considered in Table~\ref{tab:CME}.  
We take as representative the parameters for $75^\circ$ and compare these in Figure~\ref{fig:solar} to the observed parameters of the sample of solar CMEs compiled by \citet{Yashiro2009} and analysed by \citet{Drake.etal:13}.  Also shown are the parameters for the 1859 Carrington event, based on the recent analysis of \citet{Cliver.Dietrich:13} and the purported 775~AD event proposed by \citet{Melott.Thomas:12} and further analysed (and questioned) by \citet{Cliver.etal:14}.

\citet{Drake.etal:13} obtained best-fit power law relations between CME mass and kinetic energy and associated flare X-ray fluence.  The indication is that the CME mass is somewhat higher and the kinetic energy perhaps lower than the solar extrapolation.  Considering the scatter of the solar results and the fairly large Algol CME uncertainties, the results do lie in the extension of the solar CME trends.   This result is important because it is an indication that the relation between total fluence and CME mass and kinetic energy in the solar case could be extended to more active stars.

\subsection{CME dynamics}
\label{s:dynamics}

In order to examine the kinematic properties of a CME it is also critical to account for the forces that drive its motion. The forces acting on a CME are a) the Lorentz force, b) the gravitational attraction of the star, and c) the drag force due to the interaction with the stellar wind \citep{Cargill2004}. 
In the solar case, the Lorentz force dominates in the region close to Sun and leads to the CME acceleration,  whereas the drag force has the effect of converging the CME speed that has been gained out of the initial acceleration phase to the solar wind speed. So, depending on whether or when the CME is fast ($u\sim10^8$ cm~s$^{-1}$) or slow ($u\propto10^7$ cm~s$^{-1}$) the drag force will cause a gradual deceleration or acceleration, respectively \citep{Cargill2004}.

\cite{Vrsnak2004} analyzed the motion of 5000 CMEs from 2 to 30 $R_\odot$, finding an anticorrelation between CME acceleration and velocity. They concluded that there is a percentage ($\sim$14\%) of fast CMEs that accelerate and an even smaller percentage of slow CMEs ($\sim$7\%) that slightly decelerate outwards. 
Their interpretation was that the Lorentz force was responsible for those deviations, i.e.\ the Lorentz force might be non negligible at larger distances and it can have the opposite sign than initially thought, pulling the CME back towards the Sun. An alternative explanation was offered by \cite{Ruzdjak.etal:05}, who suggested that the deviation might be the result of the drag force, due to interaction of the CME with the fast wind.

The hydrogen column density decay rate suggests that we are already in the propagation phase of the CME, thus the acceleration took place before $t=31.5$~ks. 
This is consistent with the main CME acceleration seen in the solar case \citep[e.g.][]{Cargill2004,Vrsnak2004}, and indicates that acceleration has taken place close to the star shortly after the eruption.  The uncertainty in the power law decay of $N_H$ with time unfortunately precludes any investigation of subsequent more minor acceleration or deceleration.  

According to \cite{Moon.etal:02}, backed up by later analysis \citep[e.g.][]{Yashiro2009}, CMEs associated with stronger flares are faster, with a CME-flare association rate that increases with the CME speed. If the same trends are valid here like on the Sun and since we are examining a superflare, not only there is almost the certainty of an associated CME eruption, but also we expect a high CME speed.  There have been several statistical studies associating CME to flare characteristics in the solar regime \citep[e.g.][]{Moon.etal:02,Yashiro2009,Salas-Matamoros.Klein:15}. An empirical relation was reached in \cite{Salas-Matamoros.Klein:15} associating the CME speed $v_{CME}$ and the associated flare X-ray fluence, $F_X$,
\begin{eqnarray}
\log{(v_{CME})}=(0.22\pm0.05)\log{(\mathrm{F_X})}+(3.21\pm0.10) {,}
\label{e:v_cme}
\end{eqnarray}
For the Algol superflare, $F_X =3.6\times 10^6\;\mathrm{Jm^{-2}}$. If the CME investigated in this paper was a solar CME then the corresponding velocity from Equation~\ref{e:v_cme} would be of the order $v_{CME}=4.5\times 10^9$~cm~s$^{-1}$, or more than an order of magnitude larger than the largest velocities observed in solar CMEs.

A firm lower limit on the CME velocity is that corresponding to the flare size length scale derived in Section~\ref{s:speed}, $v_{CME} \geq 2.5\times 10^7$~cm~s$^{-1}$. An upper limit is difficult to determine since we have no firm constraint on the maximum size of the CME during the observations, but the speed derived assuming the dynamic length scale, $v_{CME} \approx 6.6\times 10^8$~cm~s$^{-1}$, is still an order of magnitude less than that from the extrapolation of the solar observation. The implied length scale for the solar speed extrapolation is similarly larger, and implies commensurately larger mass and kinetic energy.  For $v_{CME}=4.5\times 10^9$~cm~s$^{-1}$, the implied length scale is $S_1\sim 105\,R_{Algol,B}= 7\,S_{dyn}$, and the mass and kinetic energy are 
$M=1.0\times 10^{25}$~g and $E=1.1\times 10^{44}$~erg, respectively, for an opening angle of $75^\circ$.  As we discuss below, such high mass and energy requirements seem unlikely to be fulfilled and the indication is that the solar CME speed extrapolation does not work for the most energetic CMEs on active stars.


\subsection{Stored magnetic energy and size of starspots}
\label{s:energy}

The energy of a CME must ultimately derive from the magnetic energy stored in the corona. The question then is whether or not our estimate for the kinetic energy of the Algol event is reasonable on that basis.

\cite{Schrijver:12} used different historical indicators of geoactivity to estimate the frequency of the most energetic flares on the Sun and in combination with providing sunspot size relations for the associated active regions, they identified an upper limit of approximately $10^{34}$~erg for solar flare energies.  Extending their conclusions to more active stars, and considering {\it Kepler} white light flare observations and X-ray observations of the most energetic flares, they estimated a rough limit of about $10^{37}$~erg for Sun-like (G,K-type) stars on the main sequence.
Later, using a dimensionless 3D magnetohydrodynamic simulation of solar eruptive events, \cite{Aulanier:13} estimated a similar maximum energy that a solar flare can release as $\sim 6\times 10^{33}$ erg, i.e.\ six times larger than the most energetic event of 2003 November 4 that was ever directly recorded \citep{Schrijver:12} and within the range of stellar superflare energies \citep{Maehara.etal:12}. 

Surface magnetic fields on the most active late-type stars are known to reach and exceed kG strengths \citep[e.g.][]{Donati.Landstreet:09}.
Based on radiated X-ray energy and confinement requirements of the giant Algol flare, \citet{Schmitt1999Natur} deduced that magnetic fields of at least 500 G--1000 G must be present in the corona of Algol~B at heights of up to half a stellar radius, extending over a volume of at least $10^{33}$~cm$^3$. 
If kG fields pervaded the whole coronal volume of Algol~B to a height of $0.5R_{Algol,B}$, the total magnetic energy would be of the order of $3\times 10^{39}$~erg---one order of magnitude lower than the kinetic energy requirements for the dynamical length scale case of several  $5\times10^{40}$~erg.  The energy requirement then becomes uncomfortable given that a limited fraction of the magnetic energy would be available for CME acceleration.  The energy budget then points to a likely CME kinetic energy smaller than $10^{39}$~erg. Within the range of CME mass we have derived, we consider the energy requirements of a CME speed exceeding $10^9$~km~s$^{-1}$ entertained above in Section~\ref{s:dynamics} as unrealistically high for this particular CME.


A more likely CME energy can be estimated considering the sizes of starspots and local magnetic field strengths on active stars. 
\cite{Strassmeier:09} discusses the starspot sizes which lie in the range 0.1-10\% of the stellar surface for late-type (FGKM) stars, as observed with Doppler imaging.  Algol-type binaries present particular difficulties for assessing the surface spot and magnetic field distribution of the cooler component because the hotter companion tends to dominate the light output. Algol secondaries should be similar to the components of their cousins, the RS~CVn-type binaries, for which Zeeman-Doppler imaging has revealed that local magnetic field strengths in large spots can exceed 1~kG \citep[e.g.][]{Petit.etal:04,Rosen.etal:15}.  

A volume covering 10\%\ of the stellar surface up to a scale height of half the stellar radius amounts to $4\times 10^{33}$~cm$^3$ and the associated energy amounts to 2--$7\times 10^{39}$~erg if filled with a magnetic field of 1--2~kG.  \cite{Emslie.etal:12} studied the most energetic solar eruptive events from 1997--2003 and found that about 25\% of the stored non-potential magnetic energy in an active region gets transfered to the CME, mainly as kinetic energy.  \citet{Aulanier:13} found from a 3D MHD simulation of an eruptive flare from a highly sheared bipole that 19\%\ of the bipole energy is converted into flare energy (although only 5\%\ of this energy was converted to CME kinetic energy).  A conversion efficiency of 20\%\ would imply a maximum possible CME energy of $10^{39}$ for the Algol event.  This crude estimate can be compared with Equation~4 of \citet{Aulanier:13} derived from their simulation,
\begin{equation}
E=0.5\times 10^{32}\left(\frac{B_{max}}{10^3\; \rm G}\right)^2\left(\frac{L_{bipole}}{50\; \rm Mm}\right)^3 \; \rm erg, 
\end{equation}
where $B_{max}$ is the maximum possible bipole field, and $L_{bipole}$ the separation.  Taking $B_{max}\sim 10$~kG and a separation comparable to the stellar radius, $L_{bipole}\sim 2000$~Mm, $E\sim 3\times 10^{38}$~erg, which we take as a likely limit to the true CME kinetic energy.  This value is still an order of magnitude larger than the X-ray fluence, and within the observed scatter of ratios of solar flare and CME kinetic energies.  For our CME model, this energy corresponds to a mass of $M=2.1\times 10^{22}$~g, a length scale of $S_1=3.8R_{Algol,B}$, and a speed of $1.7\times 10^8$~cm~s$^{-1}$ which we adopt as more realistic upper limits to these quantities.

\subsection{Mass}
\label{s:cmemass}

The upper end of the mass range inferred for the CME using the dynamic length scale of $\propto 10^{23}$~g is perhaps unreasonably large from at least two different perspectives.  Firstly, unless events like the 1997 one are extremely rare, the implied mass loss rate from such CMEs becomes implausibly high, with only one event per year needed over a billion year timescale to lose up to 10\% of the stellar mass. The true frequency of these immense flares is difficult to estimate, but the fact that giant flares of similar energies have been observed on several different stars \citep[see, e.g.,][]{Favata:02} points to them being not such an uncommon phenomenon.  Secondly, the mass upper limit is much larger than the mass of the Algol corona.  Solar CMEs generally eject relatively cool plasma lifted from the chromosphere and lower corona \citep[e.g.][]{Webb:12}.  The volume emission measure of the solar corona varies with the solar cycle, but taking an average value of $n_e^2 V=10^{50}$~cm$^{-3}$ and an electron density $n_e=10^9$~cm$^{-3}$ \citep{Laming.etal:95}, the implied mass of the solar corona is about $2\times 10^{17}$~g, which is slightly larger than the most massive solar CMEs.

The total mass of the corona of Algol~B can be estimated from its ``quiescent'' coronal emission. \citet{Favata1999} found a total quiescent volume emission measure of $n_e^2 V=3\times 10^{53}$~cm$^{-3}$.  Based on high resolution {\it Chandra} spectroscopy of Algol and similar active stars, such emission arises from a range of plasma density environments, between $10^{10}$ and several $10^{12}$~cm$^{-3}$ \citep[e.g.][]{Testa.etal:04}.  Assuming approximately half originates at lower densities, the {\em emitting} coronal volume is approximately $10^{33}$~cm$^3$ and the total mass is about $2\times 10^{19}$~g. The CME material in the 1997 event must then originate from cooler, lower-lying plasma, which is not inconsistent with solar observations. 

\subsection{Other evidence for stellar CMEs}

The 1997 Algol event is not the only observational evidence for CMEs on other stars. As noted by \citet{Leitzinger.etal:14}, suspected CMEs have been highlighted from similar observations of X-ray absorption associated with large flares and from flare-associated blue shifts of Balmer lines \citep{Houdebine.etal:90,Guenther.Emerson:97,Bond.etal:01,Fuhrmeister.Schmitt:04,Leitzinger.etal:11,Vida.etal:16}.  While the 1997 Algol event remains by far the best example, the absorption signatures noted by \citet{Haisch.etal:83,Ottmann.Schmitt:96,Tsuboi.etal:98,Franciosini.etal:01,Pandey.Singh:12} could potentially be used to estimate useful CME parameters such as has been done here.  The parameters of CMEs inferred from blue-shifted spectral lines are somewhat more difficult to assess. Velocities estimated from blue shifts contain large uncertainties due to projection effects \citep[e.g.][]{Leitzinger.etal:11} and often lie in the local plasma flow range, i.e.\ a few tens to about 100$\mathrm{km\ s^{-1}}$ \citep{Bond.etal:01,Fuhrmeister.Schmitt:04,Leitzinger.etal:11}.  This makes them difficult to distinguish from smaller scale events, such as chromospheric brightenings \citep{Kirk.etal:17} or chromospheric evaporation \citep{Teriaca.etal:03}. \citet{Leitzinger.etal:14} reach the conclusion that the CME flux or mass is the main parameter that controls the detection efficiency of the Doppler-shift method. Given their importance, further examination of stellar CME candidate parameters would be worthwhile. 

\section{Conclusions}

The 1997 Algol superflare with associated absorption temporal profile observed by \cite{Favata1999} is arguably the best candidate for a CME detected in another stellar system.  While a CME is not the only possible explanation for the absorption, all clues point in that direction and this scenario is reinforced by the absorption decay agreeing very well with an inverse square law decline compatible with a quasi-uniform expansion.

After choosing physically inspired length scales, namely a flare size and a dynamic one, we were able to estimate lower and upper limits for the CME speed, mass and kinetic energy using the ice-cream cone model commonly applied to solar CMEs. While our lower limits are firm, the upper limits are characterized by large uncertainties, as they derive from drawing a  parallel between a CME acceleration length scale on Algol B and that of solar CMEs. By estimating the maximum stored magnetic energy in a starspot, we are able to place further more stringent constraints on our upper limits.  We find the likely CME mass and kinetic energy to have been in the ranges $2\times 10^{21}$--$2\times 10^{22}$~g and 
$7\times 10^{35}$--$3\times 10^{38}$~erg, respectively. 

The results are in reasonable agreement with relations between CME mass and kinetic energy and the X-ray fluence of the associated flare 
revealed by statistical studies in the solar regime \citep{Yashiro2009,Drake.etal:13} when extrapolated to the extreme energies of the Algol event. We find the Algol CME to have a likely mass lying a little higher and a kinetic energy a little lower than the extended trends. The general agreement with these trends is an indication that even in much more active stars than the Sun, such as Algol B, similar fundamental processes drive transient phenomena linked to the underlying stellar magnetic fields.  If universal, such relations would represent a breakthrough in the ability to infer CME activity on stars that cannot otherwise be easily detected. 

The Algol flare and CME are extreme phenomena, probably marking the upper limits of stellar activity events. We underline the importance of exploring further the CME--flare relation in other active stars that are expected to populate the region between the solar and the Algol event studied here.





\acknowledgments
SPM was supported by NASA Living with a Star grant number NNX16AC11G. JJD was funded by NASA contract NAS8-03060 to the {\it Chandra X-ray Center} and thanks the Director, Belinda Wilkes, for continuing advice and support. OC was supported by NASA Astrobiology Institute grant NNX15AE05G. JDAG was supported by {\it Chandra} grants AR4-15000X and GO5-16021X. CG was supported by SI Grand Challenges grant ``Lessons from Mars: Are Habitable Atmospheres on Planets around M Dwarfs Viable?''. 


\bibliographystyle{yahapj}
\bibliography{references}


\end{document}